\documentstyle[prl,aps,epsf,psfig]{revtex}
\begin{document}
\twocolumn[\hsize\textwidth\columnwidth\hsize
           \csname @twocolumnfalse\endcsname
\title{Surface charge of a flat superconducting slab in the
Meissner state}
%================================================================
\author{Jan Kol\'a\v{c}ek, Pavel Lipavsk\'y}
\address{Institute of Physics ASCR,
         Cukrovarnick\'a 10, 16253 Prague 6,Czech Republic},
\author{Ernst Helmut Brandt}
\address{Max-Planck-Institut f\"ur
             Metallforschung, D-70506 Stuttgart, Germany}
\maketitle
\begin{abstract}
%================================================================
The electrostatic potential in the flat superconducting slab is
evaluated in the framework of the Ginzburg-Landau theory extended
by Bardeen to low temperatures. For magnetic fields below
$B_{c1}$, we discuss the formation of a surface charge induced
by the Bernoulli potential of the suppercurrents.
\end{abstract}
\vskip2pc]

\section{Introduction}
%==============================================================
A diamagnetic current in a superconductor creates an electrostatic
potential known as the Bernoulli potential \cite{37Bopp,50London}.
Early studies based either on the electrodynamics of the ideal
charged fluid \cite{37Bopp}, on the London approximation
\cite{50London} or on the local approximation of the BCS theory
\cite{68Adkins,69Rickayzen,75Hong} provide the electrostatic
potential as a function of the condensate velocity. This approach
is not applicable to strongly inhomogoneous systems like Abrikosov
vortices or surfaces where non-local quantum corrections are
inevitable.

A simple non-local theory of the electrostatic potential in stationary
superconductors has been introduced in \cite{KLB01}. It is based on
the Ginzburg-Landau (GL) theory and neglects the so called
quasiparticle screening of van~Vijfeijken and Staas
\cite{64vanVijfeiken} and the thermodynamic correction predicted
by Adkins and Waldram \cite{68Adkins} and evaluated by Rickayzen
\cite{69Rickayzen}.

A theory of Ginzburg-Landau type which covers the quasiparticle
screening and the thermodynamic corrections, has been introduced in
\cite{01KoLi}. Its implementation to the Abrikosov vortex lattice
is in preparation. In this paper we employ a non-local approach
to discuss the formation of the surface charge. We study the simple
system of a superconducting flat slab in a parallel magnetic
field below $B_{c1}$.

The formation of a surface charge induced by the Bernoulli potential
has been already studied by Jakeman and Pike \cite{JP67}. Starting
from the time-dependent Ginzburg-Landau theory, they have derived a
Poisson equation for the electrostatic potential with screening
over the Thomas-Fermi length and with a source term that includes the
quasiparticle screening of van~Vijfeijken and Staas
\cite{64vanVijfeiken}. Except for the Thomas-Fermi screening their
approach is local, what implies that a surface charge is formed on
the scale of the Thomas-Fermi screening length. Here we show that the
non-local corrections to the electrostatic potential are dominant over
the Thomas-Fermi screening so that the surface charge forms on the
scale of the GL coherence length.

\section{Extended Ginzburg-Landau theory}
%================================================================
Our treatment of the electrostatic potential starts from the free
energy
\begin{equation}
{\cal F}=\int d{\bf r}\left(f_{\rm con}+f_{\rm kin}+
f_{\rm mag}\right)+{\cal F}_{\rm Coul}.
\label{F1}
\end{equation}
The kinetic energy of the condensate is of the GL form \cite{GL50}
\begin{equation}
f_{\rm kin}={1\over 2m^*}\left|(-i\hbar\nabla-e^*{\bf A})
\psi\right|^2,
\label{F2}
\end{equation}
where $\psi$ is the GL wave function, $m^*=2m$ and $e^*=2e$ are
mass and charge of a cooperon. The condensation energy follows
the two-fluid model of Gorter and Casimir \cite{34Gorter},
\begin{equation}
f_{\rm con}=U-\varepsilon_{\rm con} {2\over n}\left|\psi\right|^2-
{1\over 2}\gamma T^2\sqrt{1-{2\over n}\left|\psi\right|^2}
\label{F3}
\end{equation}
with the condensation energy per volume given by Sommerfeld's
$\gamma$ as $\varepsilon_{\rm con}={1\over 4}\gamma T_c^2$. The
total density is the sum of the condensate and normal electrons,
$n=2|\psi|^2+n_n$. The internal energy, $U$, the linear coefficient
of the specific heat per volume, $\gamma$, and the critical
temperature, $T_c$, are functions of the density $n$.

The magnetic free energy,
\begin{equation}
f_{\rm mag}=-{1\over2\mu_0}\left({\bf B}-{\bf B}_a\right)^2,
\label{F4}
\end{equation}
depends on the applied magnetic field ${\rm B}_a$ and the
internal field ${\bf B}=\nabla\times{\bf A}$. The Coulomb energy
is treated in the non-relativistic approximation,
\begin{equation}
{\cal F}_{\rm Coul}={1\over 2}\int d{\bf r}d{\bf r'}
{e^2\over 4\pi\epsilon}{1\over|{\bf r}-{\bf r'}|}
\rho({\bf r})\rho({\bf r'}),
\label{F5}
\end{equation}
where $\rho=e^*|\psi|^2+en_n+\rho_{\rm latt}$ is the charge
deviation from neutrality. We do not assume any external
electric field.

Except for the Coulomb potential, this free energy has been
introduced by Bardeen \cite{55Bardeen} soon after the paper
of Ginzburg and Landau \cite{GL50} as its simple extension to
low temperatures.

In the stationary state the free energy has a minimum which
can be found by the variation principle \cite{W96}. Variations
with respect to $\bar\psi$, ${\bf A}$ and $n_n$ lead to a set of
extended Ginzburg-Landau equations:\\
the Schr\"odinger equation,
\begin{eqnarray}  \label{extGL}
%                 ================================
&&{1\over 2m^*}\left(-i\hbar\nabla-e^*{\bf A}\right)^2\psi
\nonumber \\
&&+\left(-{2\over n}\epsilon_{\rm con}
     +{\gamma T^2\over 2n}{1\over\sqrt{1-{2\over n}|\psi|^2}}
   \right)\psi=0 ,
\end{eqnarray}
the Maxwell equation,
\begin{eqnarray}  \label{Max1}
%                 ==========================
\nabla\times\nabla\times{\bf A}
 = \mu_0{e^*\over m^*}{\rm Re}\bar\psi
     \left(-i\hbar\nabla-e^*{\bf A}\right)\psi.
\end{eqnarray}
and the screened Poisson equation,
\begin{eqnarray}  \label{3GL}
%                 ================================
e\varphi&-&\lambda_{TF}^2\nabla^2e\varphi
\nonumber\\
&=&-{\partial\epsilon_{\rm con} \over \partial n}
       {2\over n} \left|\psi\right|^2
  -{1\over 2} T^2{\partial \gamma \over \partial n}
      \sqrt{1-{2\over n}\left|\psi\right|^2}
\nonumber\\
&&-{1\over 2nm^*}\bar\psi\left(-i\hbar\nabla-e^*{\bf A}
\right)^2\psi.
\end{eqnarray}
We note that variations result directly in a condition for
the charge density. The electrostatic potential $\varphi$ has been
introduced to simplify expressions. It links to the charge density
via the plain Poisson equation
\begin{eqnarray}  \label{Max2}
%                 ================================
-\varepsilon\nabla^2\varphi=e^*|\psi|^2+en_n+\rho_{\rm latt}.
\end{eqnarray}

The last term on the right hand side of (\ref{3GL}) is the
non-local Bernoulli potential \cite{KLB01}. It is proportional
to $|\psi|^2/n$ in the spirit of the quasiparticle
screening \cite{64vanVijfeiken}. The two other terms combine
into the thermodynamic correction introduced by Rickayzen
\cite{69Rickayzen}. Note that the second term is the
thermoelectric field of the normal metal reduced by a factor
$\sqrt{1-2|\psi|^2/n}$. The screening on the Thomas-Fermi length
is covered by the differential term in the left hand side of
(\ref{3GL}).

The usual GL theory has two independent variables, ${\bf A},
\psi$ and consequently only two differential equations have to be
solved. Including the electrostatic potential we have four
indeterminates ${\bf A}, \psi, \varphi, n_n$ so that the set of
four coupled equations (\ref{extGL} - \ref{Max2}) should be solved.
In reality the problem is not so complicated. The second GL
equation (\ref{Max1}) is independent of $n_n$ and $\varphi$.
The first GL equation (\ref{extGL}) is independent of $\varphi$
while it depends on $n_n$ exclusively via $n=n_n+2|\psi|^2$.
Since $n$ is very large in metals and charge perturbations are
rather small, one can neglect deviations from the charge
neutrality in the first GL equation. In this approximation,
equations (\ref{extGL},\ref{Max1}) are solved separately. Once
${\bf A}$ and $\psi$ are known, equation (\ref{3GL}) provides us
with the electrostatic potential and the charge is found from the
Poisson equation (\ref {Max2}).

\section{Flat slab in magnetic field}
%================================================================
Let us demonstrate the behavior of the electrostatic potential
in a superconducting slab in parallel magnetic field.
We assume a superconductor of type II in a magnetic field
below $B_{c1}$, i.e., in the Meissner state with the magnetic
field penetrating from the surface.

We take the direction $x$ perpendicular to the slab limited to
the interval $(-d,d)$. The magnetic field we choose in the direction
$z$, ${\bf B}\equiv (0,0,B)$, and the vector potential points in the
direction $y$, ${\bf A} = (0,A,0)$. All functions depend
exclusively on the coordinate $x$ and the wave function is real.

We rescale all functions into their dimensionless counterparts,
$x=\tilde x\lambda_0$, $A=\tilde A\Phi_0/(2\pi\lambda_0)$,
$\psi=\tilde \psi\sqrt{n/2}$, $e\varphi=\tilde \varphi\hbar^2/
(4m\lambda_0^2)$, where the London penetration depth at zero
temperature is given by $\lambda_0^2=m/(e^2n\mu_0)$. Note that our
scale does not change with the temperature. In the expressions below
we skip tildes denoting new functions. Equations
(\ref{extGL},\ref{Max1}) now read
\begin{eqnarray}
%                 ================================
{\partial^2 A\over\partial x^2} - A\psi^2 = 0,
\label{Max1_fs}
\end{eqnarray}
\begin{eqnarray}
%                 ================================
{\partial^2 \psi\over\partial x^2}- A^2\psi+S
   \left(1-{t^2\over \sqrt{1-\psi^2}}\right)\psi = 0,
\label{extGL_fs}
\end{eqnarray}
with $S=\kappa^2(1-t^2)^2(1+t^2)$. As usual in the GL theory,
the only material parameter relevant after rescaling is the GL
parameter $\kappa$ defined at $T_c$. In the Bardeen extension
the rescaled equations also depend on the reduced temperature
$t=T/T_c$.

As the boundary condition we use that the vector potential is
anti-symmetric, $A(-x)=-A(x)$, with the value of the derivative
at the surface given by the applied magnetic field $B_a$. The
wave function has a zero derivative at the surface.

The third GL equation (\ref{3GL}) in the dimensionless units,
\begin{eqnarray}  \label{3GL_fs}
%                 ================================
\varphi-\tau^2{\partial^2\varphi\over\partial x^2}&=&
C_1\psi^2+2C_2t^2\sqrt{1-\psi^2}
     \nonumber\\
&&-\left(1-{t^2\over\sqrt{1-\psi^2}}\right)\psi^2 ,
\end{eqnarray}
depends on two material parameters,
\begin{eqnarray}
C_1&=&{\partial\ln\varepsilon_{\rm con}\over\partial\ln n},
\label{C1}\\
C_2&=&{\partial\ln\gamma\over\partial\ln n}.
\label{C2}
\end{eqnarray}
For Niobium these parameters can be derived from the McMillan
formula \cite{KSK74} and chemical trends \cite{VD76} giving
values $C_1=1.9$ and $C_2=0.42$.

The reduced Thomas-Fermi screening length,
\begin{equation}
\tau={\lambda_{TF}\over\lambda_0},
\label{redThom}
\end{equation}
is very small. For Niobium $\lambda_{TF}=0.7$~\AA\ and
$\lambda_0=390$~\AA . The Thomas-Fermi screening thus enter
equation (\ref{3GL_fs}) with factor $\tau^2=3\cdot 10^{-6}$.
With accuracy of the order of $10^{-5}$, the solution of
(\ref{3GL_fs}) is $\varphi=\varphi_{\rm in}+
\varphi_{\rm free}$, with the induced term,
\begin{eqnarray}  \label{3GL_in}
%                 ================================
\varphi_{\rm in}&=&C_1\psi^2+2C_2t^2\sqrt{1-\psi^2}
     \nonumber\\
&&-\left(1-{t^2\over\sqrt{1-\psi^2}}\right)\psi^2 ,
\end{eqnarray}
and the free solution
\begin{equation}
\varphi_{\rm free}=\varphi_0\cosh(x/\tau).
\label{freephi}
\end{equation}
The free solution decreases near the surface on the scale of the
Thomas-Fermi screening length $\tau=1.8\cdot 10^{-3}$.

The amplitude of the free solution has to be selected so that
the slab remains charge neutral. In other words, the free solution
supplies the surface charge which is localized on the Thomas-Fermi
screening length. The neutrality of the slab is equivalent to the
condition that the electric field vanishes at the surface. The
amplitude of the free solution is thus given by the condition
$\partial\varphi/\partial x=0$ on the surface.

Following \cite{YBPK01} it is possible to show that the free
solution is zero. The induced potential $\varphi_{\rm in}$ has
zero derivative on the surface as it follows from
(\ref{3GL_in}) and the GL boundary condition,
$\partial\psi/\partial x=0$. Accordingly, in the non-local
approach one has $\varphi_0=0$.

\section{Local versus non-local picture}
The local approximation corresponds to a neglect of the gradients
in the GL equation (\ref{extGL_fs}). The density of the condensate,
$\psi^2$, is thus found from a simple local condition,
\begin{equation}
1-{t^2\over \sqrt{1-\psi^2}} = {A^2\over S},
\label{extGL_loc}
\end{equation}
From (\ref{extGL_loc}) one obtains $\psi^2$ as a function of
the vector potential $A$. This function is first used in
(\ref{Max1_fs}) to solve for $A$ and finally substituted into
(\ref{3GL_fs}) to find the electrostatic potential $\varphi$.

The wave function $\psi$ found within the local approximation
(\ref{extGL_loc}) does not satisfy the GL boundary condition
$\partial\psi/\partial x=0$. Accordingly, the derivative of the
induced potential is non-zero, $\varphi_{\rm in}\ne 0$, so that
the free solution has to be supplemented to maintain the charge
neutrality.

Now we compare the full solution of (\ref{Max1_fs} - \ref{3GL_fs})
with the local approximation. In our treatment we assume Niobium
with Oxygen impurities of concentration giving $\kappa=1.5$. The
width of the slab we take $6\lambda_0$ so that the magnetic field
is screened from the bulk at low temperatures but penetrates the
whole slab close to $T_c$.

At low temperatures one finds that the magnetic field suppresses
the condensate only in a negligible fraction. Since the condensate
density is nearly constant, it follows from (\ref{Max1_fs}) that
the magnetic field differs only negligibly from the London
approximation, $B=B_0\cosh(x/\lambda)$, with $B_0=B_a/\cosh(d/
\lambda)$ and $\lambda\approx 1$. Of course, since corrections
beyond the London approximation are negligible, the local and
non-local approaches yield very similar magnetic fields.
\begin{figure}[h]
\label{f1}
\centerline{\parbox[c]{8cm}{
\psfig{figure=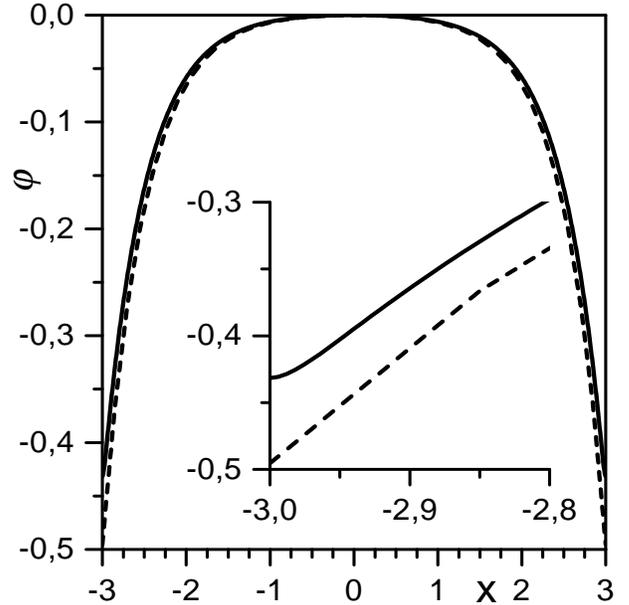,width=8cm,height=8cm}}}
\vspace{6mm}
\caption{The electrostatic potential at $t=0.1$ and $B\approx
B_{c1}$. The local approximation (dashed line) differs from the
full non-local solution (full line) mainly close to the surface
shown in the insert.}
\end{figure}

In figure~1 we show the electrostatic potential. The
local approximation agrees with the non-local approach very well
except for the close vicinity of the surface. This region is
affected by the GL boundary condition. The presented result is
for a magnetic field close to $B_{c1}$.

\begin{figure}[h]
\label{f2}
\centerline{\parbox[c]{8cm}{
\psfig{figure=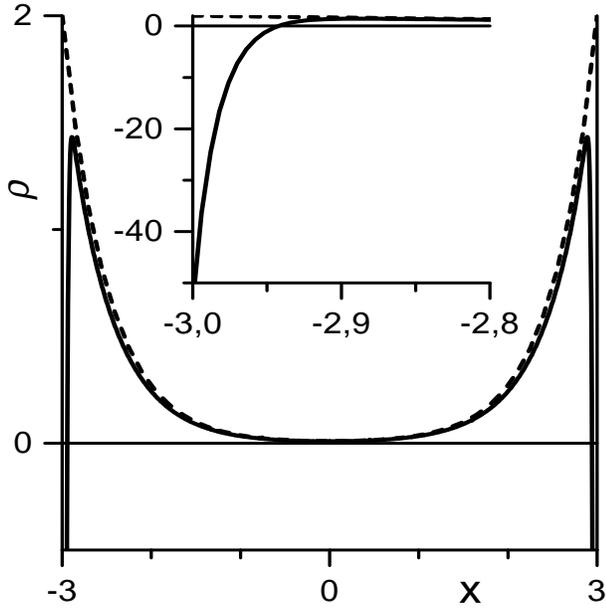,width=8cm,height=8cm}}}
\vspace{6mm}
\caption{The charge distribution corresponding to the electrostatic
potential from Fig.~1.}
\end{figure}

The formation of the surface charge is shown in Fig.~2.
The depleted charge appears in the surface layer of characteristic
width which is small compared to the London penetration depth or
the GL coherence length, however still very large on the scale of
the Thomas-Fermi screening.

\begin{figure}
\label{f3}
\centerline{\parbox[c]{8cm}{
\psfig{figure=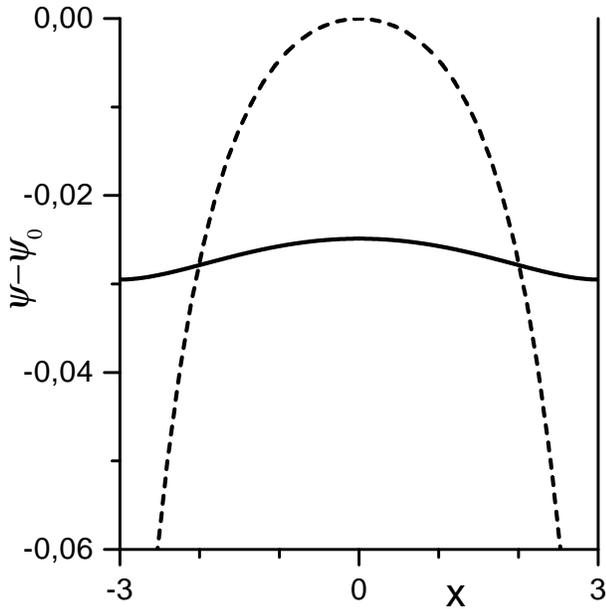,width=8cm,height=8cm}}}
\vspace{6mm}
\caption{Deviation of the wave function $\psi$ from its
equilibrium value, $\psi_0=\sqrt{1-t^4}\approx 0.6$ for $t=0.9$.
Already for a weak field, $B=0.1\,B_{c1}$, the local
approximation (dashed line) is very different from the full
non-local solution (full line).}
\end{figure}

For temperatures close to $T_c$ the local approximation becomes
unreliable if one looks for the electrostatic potential or the
related charge distribution. This contrasts with the magnetic
properties. For temperature $t=0.9$ numerical results confirm
that for a weak applied field, $B=0.1\,B_{c1}$, the magnetic
field is quite well described by the London approximation,
$B=B_0\cosh(x/\lambda)$, with $B_0=B_a/\cosh(d/\lambda)$ and
$\lambda=1.7$. It again follows from a small effect of the
magnetic field on the wave function which is nearly constant
across the slab and reduced by 5\% compared to the bulk value
$\psi_0=\sqrt{1-t^4}=0.6$, see Fig.~3. One can see that
the local approximation for the deviation of the wave function
is rather bad in this case. Indeed, the coherence length is
comparable to the width of the slab so that the GL boundary
condition is essential in the whole volume.

\begin{figure}
\label{f4}
\centerline{\parbox[c]{8cm}{
\psfig{figure=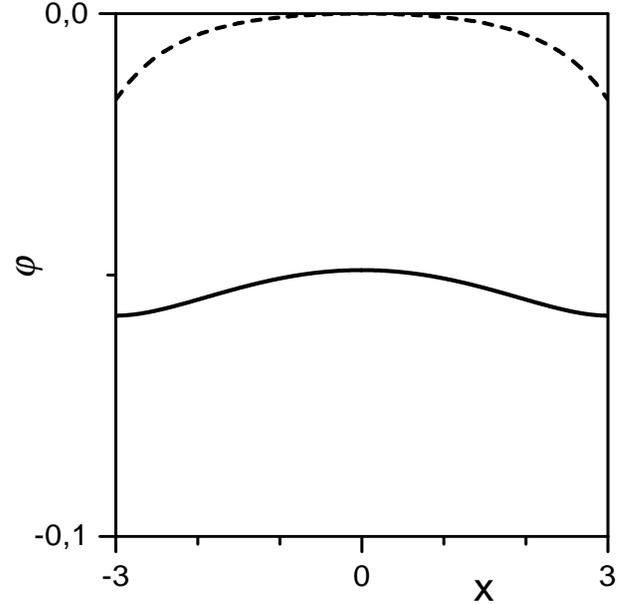,width=8cm,height=8cm}}}
\vspace{6mm}
\caption{The electrostatic potential
$\varphi=\varphi_{\rm in}+ \varphi_{\rm free}$ for $t=0.9$ and
$B=0.1\,B_{c1}$. The constant term is chosen so that both potentials 
(16) and (17) would reach zero in the bulk of a thick sample.}
\end{figure}

\begin{figure}
\label{f5}
\centerline{\parbox[c]{8cm}{
\psfig{figure=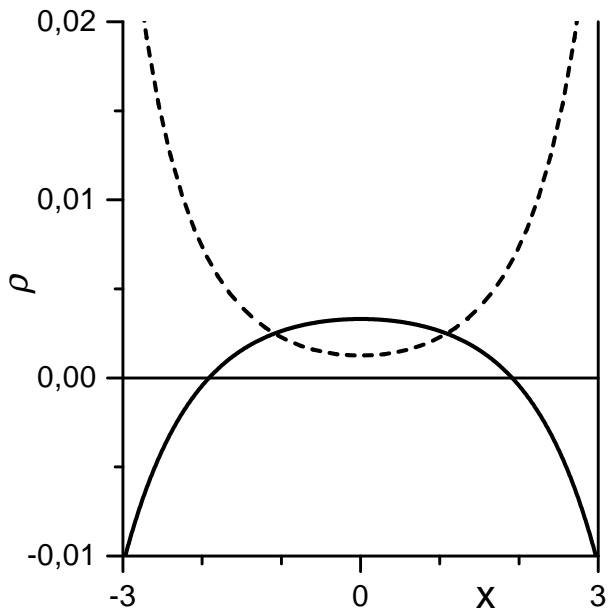,width=8cm,height=8cm}}}
\vspace{6mm}
\caption{The charge distribution corresponding to the electrostatic
potential from Fig.~4.}
\end{figure}

The local and non-local values of the electrostatic potential
shown in Fig.~4 differ appreciably. First, the non-local
potential is shifted down compared to the local approximation. We
note that in both approximations the initial is identified with
the bulk of a thick sample. The constant shift, however, is not
important for the distribution of the charge shown in
Fig.~5. One can see that in the local approximation
the charge is positive everywhere. The charge neutrality is
maintained by an invisibly thin depleted layer on the scale
$\tau\approx 10^{-3}$. In the non-local approach, the layer of
the depleted charge extends over the region comparable with
the GL coherence length.

In conclusion, the formation of the surface charge due to the
electrostatic potential caused by diamagnetic currents has been
treated within the local and non-local approaches. While the
local approach requires to include the surface on the scale of
the Thomas-Fermi screening length, the non-local approach predicts
that the surface charge extends over a layer of width comparable
with the coherence length.

\vspace{5mm}\noindent
This work was supported by GA\v{C}R 202000643, GAAV A1010806 and
A1010919 grants. The European ESF program VORTEX is also gratefully
acknowledged.
\vspace{-5mm}
% References
%================

\end{document}